\documentclass[final]{svjour2}                    % onecolumn
\smartqed  % flush right qed marks, e.g. at end of proof
\usepackage{graphicx}
\usepackage[utf8x]{inputenc}
\usepackage{amsmath}
\usepackage{amssymb}
\usepackage{sidecap}
\usepackage{caption}
\newcommand{\KPperiod}{a+b}
\newcommand{\vzero}{V_0}
\newcommand{\Tcritic}{T_c}
\newcommand{\GrandPot}{{\Omega}}

\newcommand{\Tzero}{T_0}
\newcommand{\lambdazero}{\lambda_0}
\newcommand{\bosondensity}{\eta_b}
\newcommand{\KPperiodzero}{{(\KPperiod})/\lambdazero}

\newcommand{\uzerotilde}{{\tilde{u}_0}}
\newcommand{\vzerobar}{{u_0}}
\newcommand{\ezbar}{\bar{\varepsilon}_z}

\newcommand{\energy}{\varepsilon}
\newcommand{\epart}[1]{{\energy}_{#1}}
\newcommand{\ez}[1][]{{\epart{#1z}}}
\newcommand{\ezero}{{\energy_0}}

\begin{document}

    \title{Universal behavior of the BEC critical temperature for a multislab ideal Bose gas}
    %\subtitle{Do you have a subtitle?\\ If so, write it here}

    %\titlerunning{Short form of title}        % if too long for running head

    \author{O. A. Rodr\'{i}guez \and M. A. Sol\'{i}s}

    %\authorrunning{Short form of author list} % if too long for running head

    \institute{
    	O. A. Rodríguez López \at Posgrado en Ciencias Físicas, UNAM, \email{oarodriguez@fisica.unam.mx}
        \and
        M. A. Solís \at Instituto de Física, UNAM, \email{masolis@fisica.unam.mx}
    }

    \date{Received: date / Accepted: date}
    % The correct dates will be entered by the editor

    \maketitle

    \begin{abstract}
    	For an ideal Bose-gas within a multi-slab periodic structure, we discuss the effect of the 
    	spatial distribution of the gas on its Bose-Einstein condensation critical temperature 
    	$T_c$, as well as on the origin of its dimensional crossover observed in the specific heat. 
    	The multi-slabs structure is generated by applying a Kronig-Penney potential to the gas in 
    	the perpendicular direction to the slabs of width $b$ and separated by a distance $a$, and 
    	allowing the particles to move freely in the other two directions. We found that $T_c$ 
    	decreases continuously as the potential barrier height increases, becoming inversely 
    	proportional to the square root of the barrier height when it is large enough. This 
    	behavior is {\it universal} as it is independent of the width and spacing of the barriers. 
    	The specific heat at constant volume shows a crossover from 3D to 2D when the height of the 
    	potential or the barrier width increase, in addition to the well known peak related to the
    	Bose-Einstein condensation. These features are due to the trapping of the bosons by the 
    	potential barriers, and can be characterized by the energy difference between the energy 
    	bands below the potential height.

    	% \PACS{PACS code1 \and PACS code2 \and more}
    	% \subclass{MSC code1 \and MSC code2 \and more}
    \end{abstract}

    \section{Introduction}

    Quantum fluids within periodic structures have been subject of study from some decades ago.
    Important examples of this kind of systems are electron gas in layered cuprate superconductors \cite{cuprate},
    cold atomic gases in periodic structures constructed from laser beams known as optical lattices 
    \cite{opticallattice}, the well known liquid helium four or three in any dimension \cite{helium},
    electons or holes in semiconductor superlattices \cite{superlattice} or Cooper pairs in tube bundle superconductors
    \cite{tubebundle}. The physical properties of a constrained quantum gas come from the effects 
    of the external potentials and the interactions between particles. Here we are giving a 
    detailed description of the effects of the constraints on the Bose gas properties before to
    address a more complete description including interactions. Let us consider a 3D ideal Bose gas 
    (IBG) within a lattice composed by multiple slabs. The bosons are free to move in the $x$ an 
    $y$ directions. The lattice is modeled through a Kronig-Penney potential in the $z$ direction, 
    with potential barriers of width $b$ separated a distance $a$, so the potential period, or
    the lattice period, is $a+b$.
    We solve the Schr\"odinger equation by separation of variables. The energy of the particles with
    mass $m$ is $\energy = \epart{x} + \epart{y} + \epart{z}$, where $\epart{x} = \hbar^2 k_x^2 /
    2m$, $\epart{y} = \hbar^2 k_y^2 / 2m$, being $ \hbar \, k_x$ and $\hbar \, k_y$ the momentum of 
    free bosons in the $x$ and $y$ directions respectively, while $\epart{z}$ is found by solving 
    the Schr\"odinger equation in the $z$ coordinate with the Kronig-Penney potential
    \begin{equation}
        \label{eq:kp-potential}
        \mathcal{U}(z) = \vzero \sum_{n=-\infty}^{\infty} \Theta[z - (n-1)(a+b) - a] \,
        \Theta[n(a+b) - z],
    \end{equation}
    where $\Theta(z)$ is the Heaviside step function and $V_0$ is the height of the potential barriers. The allowed energies for a particle in the $z$-direction are given by \cite{bib:kp}
    \begin{equation}
        \label{eq:kp-dispertion-relation}
        \frac{\vzero - 2\ez}{2 \sqrt{\ez (\vzero - \ez)}} \sinh(\kappa b) \sin(\alpha a) +
        \cosh(\kappa b)\cos(\alpha a) = \cos[k_z (a+b)],
    \end{equation}
    where $\hbar\kappa = \sqrt{2m(\vzero - \ez)}$ and $\hbar \alpha = \sqrt{2m \ez}$.

    The energy spectrum consist of a series of allowed energy bands separated by forbidden
    regions, where the positive side of the $j$-th band extends from $k_z (\KPperiod) = (j-1)\pi$ 
    to $j \pi$ with $j=1,2,3,$... . The Fig. 1 in Ref.~\cite{bib:oarodriguez} shows the energy 
    spectrum, and the allowed energy bands, in terms of the dimensionless parameters $u_0 = ({2m
	(\KPperiod)^2}/{\hbar^2}) \vzero$, $\bar{\varepsilon}_z = ({2m (\KPperiod)^2}/{\hbar^2}) \ez$ 
	and $r = {b}/{a}$, for $r=1$. Pi squared times our energy unit $\hbar^2/2m(a+b)^2$  
    is equivalent to the recoil energy of an 1D optical lattice formed by the interference of two 
    counter-propagating light beams of wavelength equal to $2(a+b)$. The parameter $u_0$ represents 
    the lattice height (or depth) in our energy unit, and the quotient $r$ indicates how wide are 
    the barriers with respect to the lattice period. It also indicates if a lattice is square 
    ($r = 1$) or rectangular ($r \neq 1$).

    \section{Universal behavior of the critical temperature}

    We obtain the thermodynamic properties of the Bose gas from the Grand Potential $\GrandPot = U 
    - TS - \mu N$. In order to find the Bose-Einstein critical temperature $\Tcritic$ we use the 
    number equation, $N = N_0 + N_e$, where $N_0$ is the number of bosons occupying the ground 
    state, and $N_e$ is the number of bosons distributed over excited states. At $T = \Tcritic$, 
    $\mu(\Tcritic) = \mu_0 = \ezero$ is the energy of the ground state of the system, and all the 
    bosons are distributed over the excited states, $N \approx N_e$. The number equation becomes
    \begin{equation}
        \label{eq:critical-temperature-equation}
        N = - \frac{m V}{(2\pi)^2 \hbar^2} \frac{1}{\beta_c}  \int_{-\infty}^{\infty} dk_z \, \ln(1
        - e^{-\beta_c (\ez - \mu_0)}),
    \end{equation}
    with $\beta_c = 1/k_B \Tcritic$. From the last Eq. (\ref{eq:critical-temperature-equation}) we 
    obtain the critical temperature $\Tcritic$ in an implicit way, which is numerically calculated. 
    Our system size is infinite, and it has an infinite number of bosons, but the average particle 
    density $\bosondensity = N/V$ is a constant. An infinite IBG with the same average density but 
    with no external potential has a critical temperature given by $\Tzero = ({2 \pi \hbar^2}/{m 
    k_B}) \left[ {\bosondensity}/{\zeta(3/2)} \right]^{2/3}$, where $\zeta(3/2) \simeq 2.612$ is 
	the Riemann Zeta function of argument $3/2$. It turns out that $\Tzero$ is a good reference 
	parameter for temperatures as well as the thermal wavelength $\lambdazero = h / \sqrt{2\pi m 
	k_B \Tzero}$ is a good reference parameter for lengths, in particular for the critical 
	temperature $T_c$ and the potential spatial period $\KPperiod$, respectively.
 
\begin{SCfigure}[][b!]
        \centering
        \includegraphics[width=3.4in]{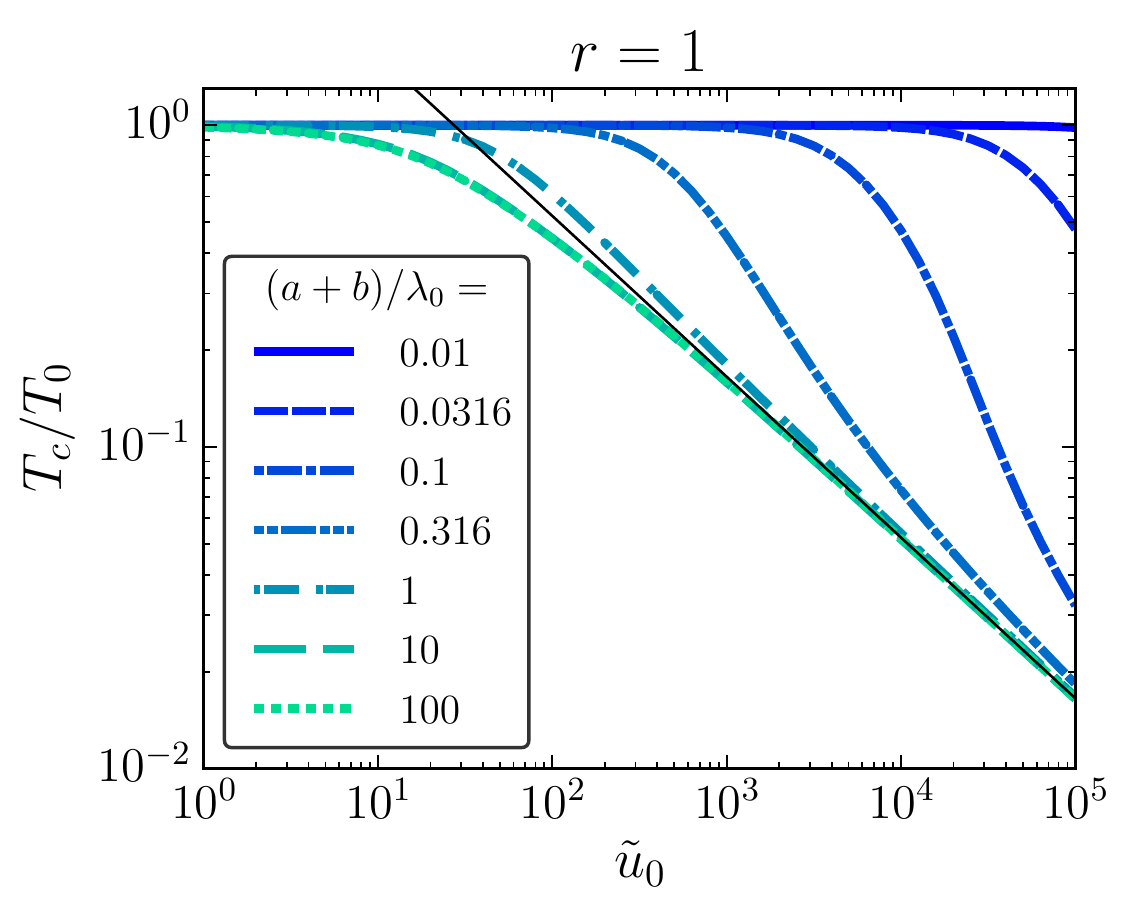}
        \caption{(Color online) $\Tcritic/\Tzero$ as a function of $\uzerotilde$. The solid, thin line
            indicates the behavior given by \eqref{eq:tc-limit-behavior-large-u0}.}
        \label{fig:tc-curvesp_r-1}
        \vspace{-0.2in}
    \end{SCfigure}

    We previously found \cite{bib:oarodriguez} that, when $u_0 \equiv (2m (a+b)^2/\hbar^2) V_0 \gg 
    1$ the critical temperature is inversely proportional to the square root of this parameter. 
    However, in order to observe separately the effects of the spatial period variation as well as 
    the increasing of the potential height on the  critical temperature, in this Section we find it 
    convenient to replace $u_0$ by $\tilde{u}_0 \equiv (2m\lambdazero^2/\hbar^2) V_0$ which is 
    period independent and it depends exclusively on changes of the potential magnitude $V_0$. The 
    behavior of the critical temperature as a function of $\tilde{u}_0$ (instead of $u_0$) is shown 
    in Fig. \ref{fig:tc-curvesp_r-1}, for $r = 1$, i.e., the width of the barriers is equal to the 
    separation between them. In the same figure each curve of $\Tcritic/\Tzero$ corresponds to a 
    particular value of the period of the potential. 
    We can see that $\Tcritic$ is a monotonically decreasing function of $\tilde{u}_0$: when the 
    magnitude of the potential barriers is small $\Tcritic$ is approximately $\Tzero$, then it 
    falls as $\tilde{u}_0$ increases, and for sufficiently high potential barriers the critical 
    temperature is proportional to $\tilde{u}_0^{-1/2}$. In addition, for $\KPperiodzero < 1$ we 
    clearly see that $T_c$ is very close to $T_0$, even for very high potential barriers. This 
    behavior is due to the fact that, when the potential period becomes very small with respect to 
    $\lambdazero$, the energy spectrum of the bosons approaches more and more to that of a free 
    particle, so $T_c$ tends to $\Tzero$. It seems as if the thin potential wells are unable to 
    capture bosons and all of them have energies above $V_0$ behaving as a free IBG. 
    
	\begin{SCfigure}[][b]
		\centering
		\includegraphics[width=3.4in]{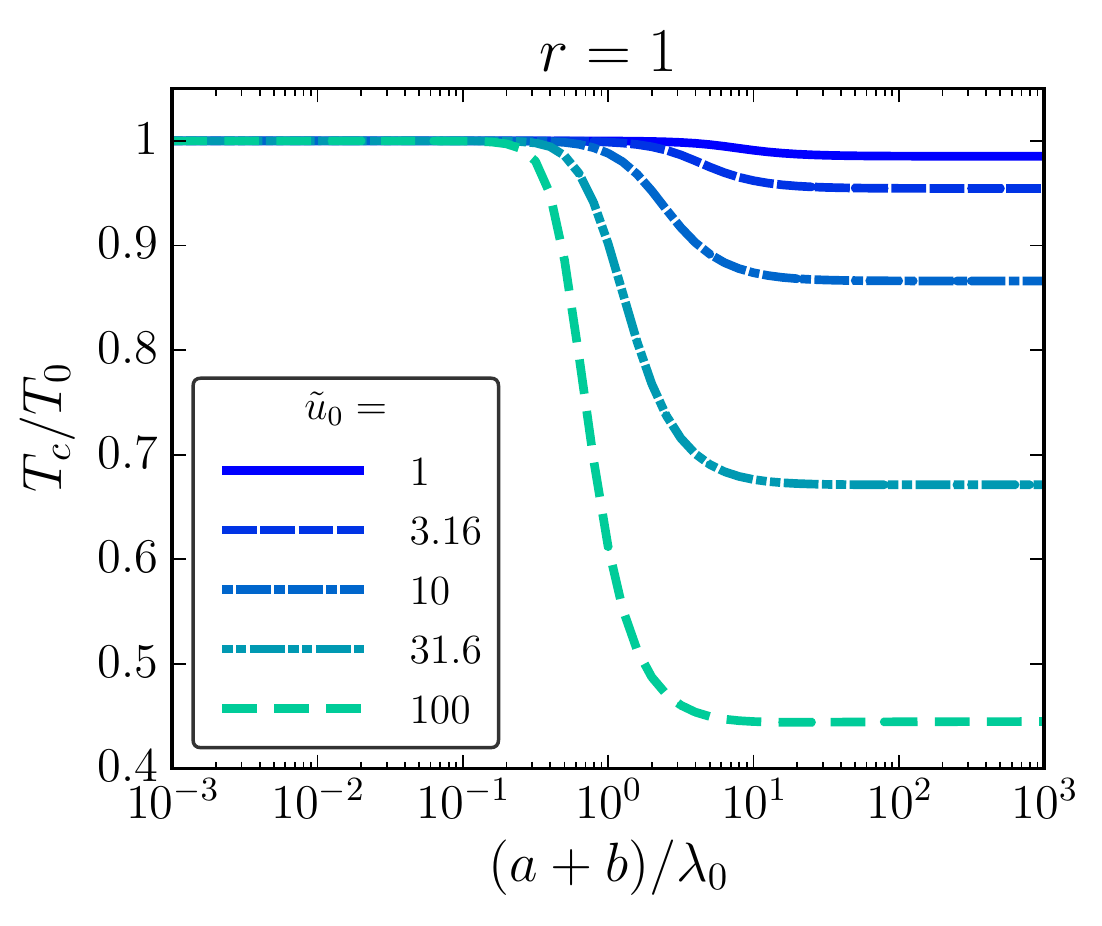}
		\caption{(Color online) $\Tcritic/\Tzero$ as a function of the potential period.}
		\label{fig:tc-curvesa0_r-1}
	\end{SCfigure}
   
    In the other hand, we found that when $[(a+b) / \lambdazero] \, r\sqrt{\tilde{u}_0} / (1+r) \gg 
    1$, the energy bands whose energies are less than $\uzerotilde$ are very narrow, and they 
    resemble the energy levels of a particle in a box of width $a$. In this case we can approximate 
    the energies of the first band through a Taylor series up to second order derivatives of 
    $\ez(k_z)$ around $k_z = 0$,
    \begin{equation}
        \ez \approx \ezero + \frac{1}{2} \, k_z^{2} \, D^{2}_{k_z} \ez \left.\right|_{k_z = 0}.
    \end{equation}
    The linear term on $k_z$ is zero because $\ez(0) = \ezero$ is a minimum of the energy. Using 
    this approximation, and taking the condition that $\KPperiodzero \gg 1$, the behavior of 
    $\Tcritic/\Tzero$ is given by the relation
    \begin{equation}
        \label{eq:tc-limit-behavior}
        \frac{r{\uzerotilde}^{1/2}}{(1+r) \zeta(3/2)} \frac{\Tcritic}{\Tzero} + \frac{1}{1+r} \left(\frac{\Tcritic}{\Tzero}\right)^{3/2} -1 = 0,
    \end{equation}
    which is potential period independent, therefore $\Tcritic/\Tzero$ depends only on 
    $\uzerotilde$ and $r$. When $r \sqrt{\uzerotilde} / (1+r) \gg 1$ the second term of 
    \eqref{eq:tc-limit-behavior} becomes negligible, and we obtain the concrete dependence of the 
    critical temperature shown in Fig. \ref{fig:tc-curvesp_r-1} as a solid thin line,
    \begin{equation}
        \label{eq:tc-limit-behavior-large-u0}
        \frac{T_c}{\Tzero} = \frac{1+r}{r} \zeta(3/2) \uzerotilde^{-1/2}.
    \end{equation}
    The numerical results plotted in Fig. \ref{fig:tc-curvesp_r-1} supports that Eq. 
    (\ref{eq:tc-limit-behavior-large-u0}) is an excellent limit relation when $\uzerotilde$ is very 
    large in comparison with $[\KPperiodzero]^{-2}$, even when $\KPperiodzero$ is smaller than 
    unity. The behavior of $\Tcritic$ as a function of the potential period can be seen in 
    Fig.~\ref{fig:tc-curvesa0_r-1}, where we note that $\Tcritic$ approximates to $\Tzero$ when the 
    potential period is small, then it falls as the period increases, and finally it goes to a 
    certain value that is $\uzerotilde$ and $r$ dependent, but becomes $\KPperiodzero$ independent. 
    This constant value is given by solving \eqref{eq:tc-limit-behavior}.
    
    It is interesting to see how the critical temperature does not necessarily approaches to $T_0$ 
    again as the potential period keeps increasing, unlike the case of a Dirac comb potential 
    \cite{bib:salas}. Even when the empty space between barriers becomes larger (as well as the 
    barriers width), the distribution of the bosons over the bands is very different to that of the 
    free Bose gas. The dependence on $\tilde{u}_0$ and $r$ that appears on the first term of 
    \eqref{eq:tc-limit-behavior} comes from the distribution of the bosons over the first energy 
    band. The dependence on $r$ in the second term of \eqref{eq:tc-limit-behavior} comes from the 
    boson distribution over the rest of the bands. Only in the case when $r = 0$ the critical 
    temperature is $T_0$. This is expected because $r=0$ means no barriers, i.e., $b = 0$, or $a 
    \to \infty$ while $b$ remains constant. In both cases it is clear why we recover the behavior 
    of the free Bose gas.
    
    \begin{SCfigure}[][b]
    	\centering
    	%\hspace*{0.3in}
    	\includegraphics[width=3.4in]{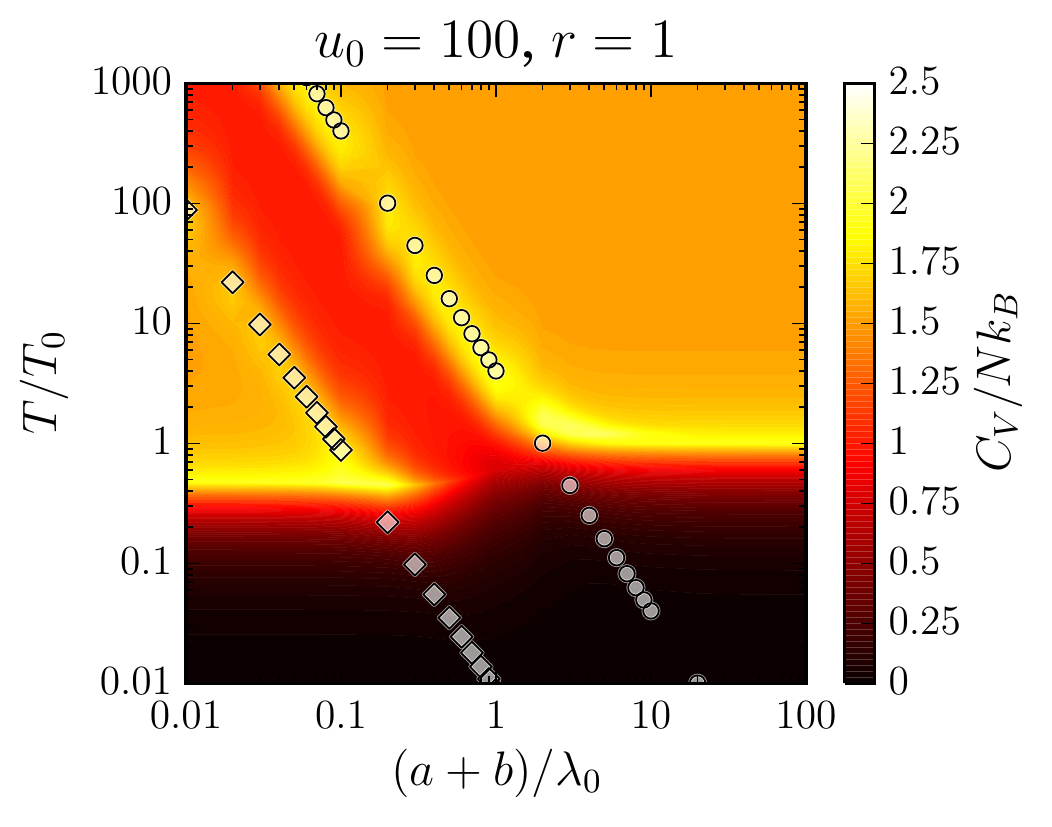}
    	\caption{(Color online) $C_V / N k_B$ contours as a function of the potential period and the temperature.}
    	\label{fig:cv-contour_u0-100_r-1}
    \end{SCfigure}

    \section{Dimensional crossover and energy spectrum}

    We have calculated the specific heat at constant volume for several combinations of the 
    parameters (see Fig. 4 of Ref. \cite{bib:oarodriguez}). The specific heat has a peak due to the 
    Bose-Einstein condensation of the confined IBG at a critical temperature $T_c \leq T_0$. 
    Above $T_c$ the specific heat has a complex structure, and for some combinations of the 
    parameters of the system it shows a dimensional crossover within an interval of temperatures 
    where it falls to a minimum approximately equal to unity; in this interval the system behaves 
    as a 2D gas, trapped by the potential in one direction but remaining free in the other two. The 
    minimum, which is a consequence of the trapping effects imposed by the potential barriers could 
    become a plateau when the potential magnitude is increased. In addition to the minimum, the 
    specific heat has two maxima. The maximum at the lower temperature marks the crossover from the 
    3D behavior of the gas to a 2D behavior. The maximum at higher temperature represents the 
    return of the system to its three-dimensional behavior or it may be seen as the superior limit 
    of the temperature interval where the crossover occurs. 
 
    In order to give a explanation about the meaning and position of the specific heat maxima in 
    terms of the energy spectrum characteristics, we realized a contour plot of $C_V/N k_B$ as a 
    function of $\KPperiodzero$ and $T/\Tzero$, for a system with ${u}_0 = 100$ and $r=1$ (see 
    Fig.~\ref{fig:cv-contour_u0-100_r-1}) where the maximum positions are highlighted with bright 
    yellow.
    
    When we associate the temperature of the maxima $T_{max}$ with a yet unknown energy chunk 
    $\Delta\ez$ characteristic of the energy spectrum such that $\Delta\ez = k_B T_{max}$, we found 
    that both lower and higher maxima temperatures follow the relation
    \begin{equation}
        \label{eq:crossover-t-edge}
        \frac{T_{max}}{\Tzero} = \frac{1}{4 \pi} \Delta\ezbar
        \left(\frac{\KPperiod}{\lambdazero}\right)^{-2},
    \end{equation}
    where $\Delta\ezbar$, the energy chunk given in recoil energy unit divided by $\pi^2$, has a 
    different value depending on whether we are describing the maximum at lower or higher 
    temperature. 
   
	\begin{SCfigure}[][b]
		\centering
		%\hspace*{0.3in}
		\includegraphics[width=3.4in]{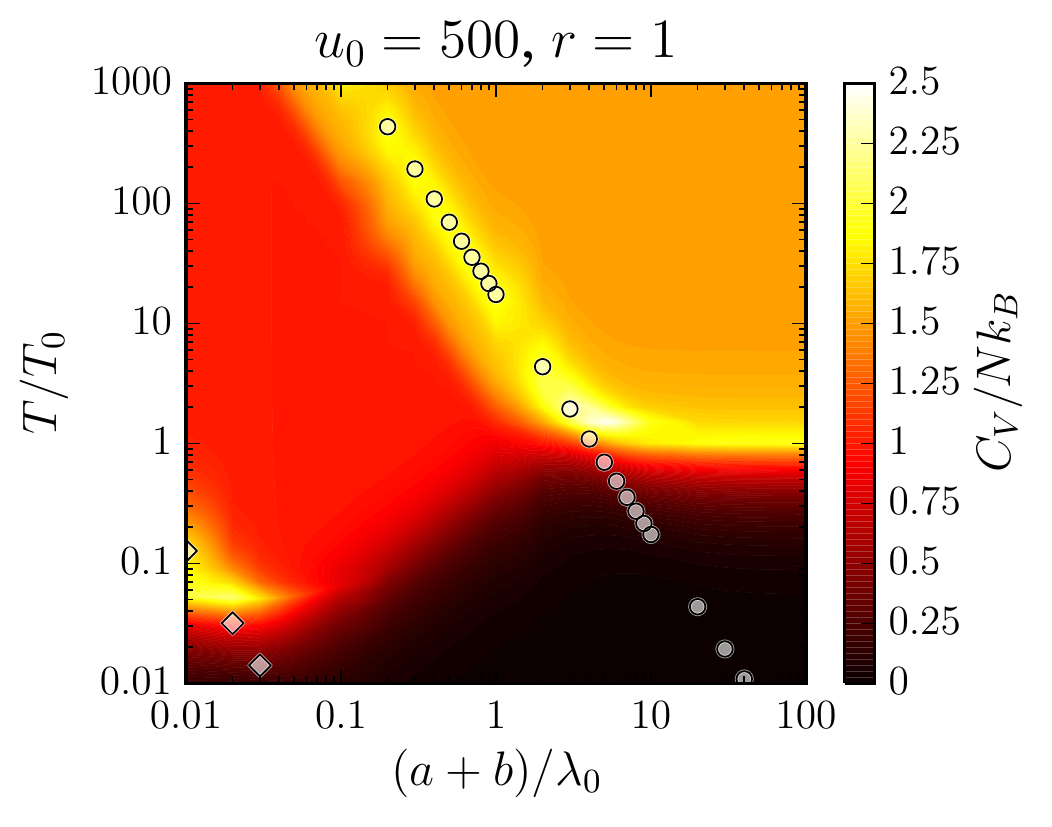}
		\caption{(Color online) $C_V / N k_B$ contours as a function of the potential period and the temperature.}
		\label{fig:cv-contour_u0-500_r-1}
	\end{SCfigure}
  
    The maximum that occurs at the higher temperature is reproduced using an energy $\Delta\ezbar$ 
    given by the difference between the bottom edge of the first energy band and the bottom edge of 
    the second energy band. In the Fig.~\ref{fig:cv-contour_u0-100_r-1} this is shown with white 
    circles. The maximum at the lower temperature occurs when $\Delta\ezbar$ is approximately equal 
    to $1/4$ of the width of the first energy band. In the Fig.~\ref{fig:cv-contour_u0-100_r-1} 
    this relation is shown with white diamonds.  
        
    As a more general result, it is found that as the magnitude of $\vzerobar$ increases the energy 
    chunk needed to reproduce the $T_{max}$ of the maximum at higher temperatures, eventually 
    corresponds to the energy difference between the bottom edges of the first and the third, 
    fourth or subsequent bands of the confined Bose gas energy spectrum. An example of this 
    behavior is show in Fig.~\ref{fig:cv-contour_u0-500_r-1} for the system with $u_0 = 500$ and 
    $r=1$, where the energy chunk $\Delta\ezbar$ associated to the maximum at the higher 
    temperature corresponds to the difference between the bottom edges of the third and first band. 
    However, the dependence of the energy chunk on $\vzerobar$ and $r$ is not trivial to find, and 
    currently we have no analytical expression to determine it. It is noticeable that, to reproduce 
    the maximum at lower temperature with expression (\ref{eq:crossover-t-edge}) we always use 
    $\Delta\ezbar$ approximately to $1/4$ of the width of the first energy band of the spectrum, 
    i.e., in this case $\Delta\ezbar$ value is potential strength independent.

    \section{Conclusions}

    In summary, we analyzed the behavior of the BEC critical temperature of an IBG confined by an 
    infinite stack of slabs as well as the effects of the external periodic potential on the 
    dimensional crossover, the specific heat maximum positions and their relation with the 
    distribution of allowed and forbidden bands in the energy spectrum. We found that the critical 
    temperature $\Tcritic$ decreases continuously as the potential barrier height increases, and it 
    becomes proportional to $\uzerotilde^{-1/2}$ when $\uzerotilde$ is much larger than 
    $[\KPperiodzero]^{-2}$. It is found that this behavior is {\it universal} as it does not 
    depends on the potential period, the barrier width nor the barrier separation. In addition, we 
    found that as the period increases the $\Tcritic$ becomes a constant  value given by 
    Eq.~\eqref{eq:tc-limit-behavior} which is spatial period independent. On the other hand, the 
    specific heat at constant volume shows a dimensional crossover within a certain interval of 
    temperatures where it behaves like a 2D gas due to the trapping by the potential barriers. The 
    region where the dimensional crossover occurs as well as the specific heat maximum positions 
    depend directly on the forbidden and allowed bands distribution in the energy spectrum. We were 
    able to find a relation between the maximum positions and characteristic energies of the energy 
    spectrum. Although the specific heat maximum position, at higher temperature, is reproduced 
    using the Eq.~\eqref{eq:crossover-t-edge} with $\Delta\ezbar$ the energy difference between the 
    bottom edges of the second and first bands, or, depending on the potential strength, the 
    difference between the bottom edges of the third and first, the bottom edges of the fourth and 
    first  and so on, we highlight that the specific heat maximum position at the lower temperature 
    is reproduced using the Eq.~\eqref{eq:crossover-t-edge} with $\Delta\ezbar$ being potential 
    strength independent and equal to $1/4$ of the width of the first band. 
 
    \vspace{0.30cm}

    \noindent {\bf Acknowledgements.}
    We acknowledge partial support from grants PAPIIT-DGAPA-UNAM IN-105011 and IN-111613,
    CONACyT 221030 and PAEP UNAM.

    % Non-BibTeX users please use
    %\begin{thebibliography}{}
    %%
    %% and use \bibitem to create references. Consult the Instructions
    %% for authors for reference list style.
    %%
    %\bibitem{RefJ}
    %% Format for Journal Reference
    %Author, Article title, Journal, Volume, page numbers (year)
    %% Format for books
    %\bibitem{RefB}
    %Author, Book title, page numbers. Publisher, place (year)
    %% etc
    %\end{thebibliography}

\end{document}